\newif\ifanonymousversion
\anonymousversionfalse
\pdfoutput=1
\documentclass[sigplan,10pt]{acmart}
\setcopyright{acmcopyright}
\renewcommand\footnotetextcopyrightpermission[1]{}
\usepackage{soul}
\usepackage{graphicx}
\newcommand{\nsf}[1]{\href{https://www.nsf.gov/awardsearch/showAward?AWD_ID=#1}{#1}}
\begin{document}
\title[Quantum Operating System Support for Quantum Trusted Execution Environments]{Quantum Operating System Support for\\ Quantum Trusted Execution Environments}
\ifanonymousversion
\author{Anonymous Authors}
\else
\author{Theodoros Trochatos}
 \affiliation{
 \institution{Yale University}
 \city{New Haven}
 \state{Connecticut}
 \country{USA}}
 \email{theodoros.trochatos@yale.edu}
\author{Jakub Szefer}
 \affiliation{
 \institution{Yale University}
 \city{New Haven}
 \state{Connecticut}
 \country{USA}}
 \email{jakub.szefer@yale.edu} 
\fi
\begin{abstract}
  With the growing reliance on cloud-based quantum computing, ensuring the confidentiality and integrity of quantum computations is paramount. Quantum Trusted Execution Environments (QTEEs) have been proposed to protect users' quantum circuits when they are submitted to remote cloud-based quantum computers. However, deployment of QTEEs necessitates a Quantum Operating Systems (QOS) that can support QTEEs hardware and operation. This work introduces the first architecture for a QOS to support and enable essential steps required for secure quantum task execution on cloud platforms.
\end{abstract}
\maketitle
\section{Introduction}
Current cloud computing platforms such as IBM Quantum \cite{ibmquantum} or Amazon Braket \cite{amazonbracket} provide users access to quantum computers. However, today, cloud providers have full visibility and control over users' submitted quantum circuits. This can expose users to the risk of proprietary quantum algorithms being reverse-engineered or stolen by the providers. Even if the cloud providers are trusted, there are other threats such as malicious insiders \cite{10288020,10545385} who can gain access to the users' circuits and steal the information.
To mitigate risks of security attacks in cloud-based quantum computation, Quantum Trusted
Execution Environments (QTEEs) have been proposed \cite{10288020,10545385,trochatos2023hardwarearchitecturequantumcomputer}. Their goal is in general to prevent the cloud provider from knowing the operation executed by the user or the results. To achieve, this QTEEs apply some sort of obfuscation on the software level, the obfuscated or protected circuit is sent to the cloud provider, who cannot understand it, and only in the trusted hardware of the quantum processor can the circuit be de-obfuscated and executed.
Currently, the missing piece of the technology is the Quantum Operating System (QOS) which can manage the user circuits and the QTEEs hardware. The QOS must manage secure loading of quantum circuits, execution, and transmission of computation results back to the users.
\section{Design of QOS Support for QTEEs}
This work outlines the mechanisms needed for securely loading quantum circuits, establishing a trusted environment for their execution, and returning the results to the users.
\subsection{Existing QTEEs}
There are already a number of QTEEs that have been proposed in the literature, which are briefly introduced below. 
\paragraph{QC-TEE} The QC-TEE work \cite{10288020} introduced the idea of adding obfuscation to quantum circuits. While digital representation of the quantum circuits can be encrypted, the circuits are eventually transformed into analog pulses before execution on quantum hardware, analog pulses cannot be encrypted -- but cloud provider can spy and attack these pulses. QC-TEE introduced hardware modifications to remove the dummy obfuscation pulses before they reach qubits. Encrypted metadata was used to allow QTEE hardware to determine which are the dummy obfuscation pulses.
\paragraph{SoteriaQ} The SoteriaQ work \cite{trochatos2023hardwarearchitecturequantumcomputer} expanded the ideas of QC-TEE \cite{10288020} and outlined detailed architecture of the circuit obfuscation. Encrypted metadata was again used to allow QTEE hardware to determine which are the dummy obfuscation pulses.
\paragraph{CASQUE} The CASQUE work \cite{10545385} introduced a new idea of extending obfuscation by swapping pulses between different control and drive channels. In the user's circuit, after transpilation, the control pulses would be swapped between different channels. On the quantum computer end, CASQUE introduced hardware modifications so that the pulses could be swapped back into the correct channels before they reach qubits. Encrypted metadata was used to allow CASQUE hardware to determine how to un-swap the control pulses.
\subsection{Life-cycle of Quantum Circuit in QTEE}
\begin{figure*}[t]
     \centering
     \includegraphics[width=0.99\linewidth]{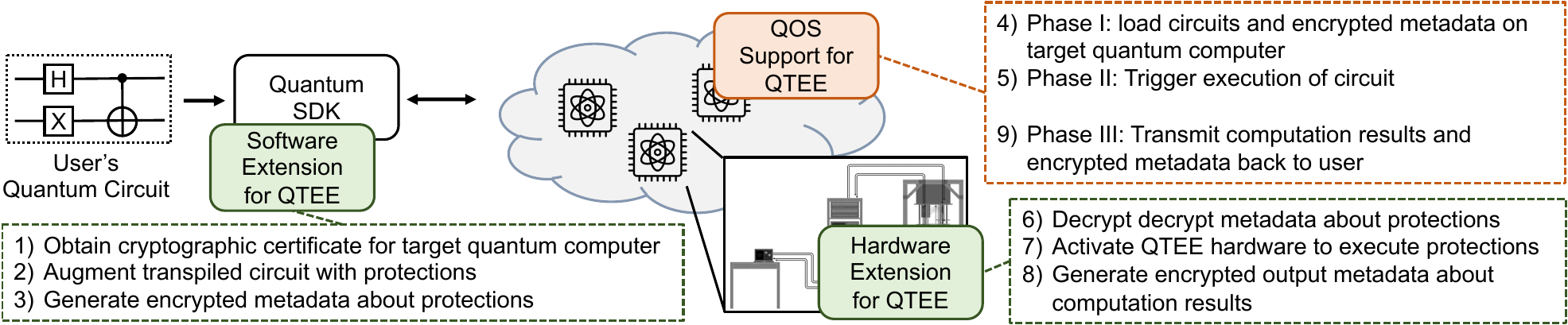}
        \caption{\small Lifecycle of quantum circuits and QOS support needed for QTEEs.}
        \label{fig_qc_tee_lifecycle_qos}
\end{figure*}

Regardless of the QTEE type, the lifecycle of a circuit in a QTEE follows three phases: I) secure loading of quantum circuits, II) execution on the quantum computing hardware, and III) transmission of computation results back to the users. Figure~\ref{fig_qc_tee_lifecycle_qos} outlines the life-cycle of a quantum circuit as handled by QOS.
\paragraph{Phase I: Secure Loading of Quantum Circuits}
On the user-end, the quantum circuit is obfuscated according to the target QTEE, and encrypted metadata is attached to the circuit. The obfuscated circuit and encrypted metadata are securely sent to the cloud provider, by an encrypted network connection. Upon decryption of the network packets, the circuit and encrypted metadata, needs to be safely stored by QOS while awaiting execution.
{\em The QOS needs to support classical, secure networking to receive users' circuits and their encrypted metadata. The QOS needs to track of the circuit and encrypted metadata once received. When storing them on the cloud, the circuit and encrypted metadata need to be associated with each other. Since the obfuscation method and encrypted metadata is specific to a particular quantum computer, the QOS scheduling also needs to be augmented to keep track of which quantum computer the circuit can execute on.}
\paragraph{Phase II: Secure Execution of Quantum Circuits}
When the circuit is ready to execute, the transpiled circuit is loaded onto the quantum controller, which, for example, in case of superconducting qubit quantum computer, generates the analog pulses that drive the qubits. In case of QC-TEE \cite{10288020}, SoteriaQ \cite{trochatos2023hardwarearchitecturequantumcomputer}, and CASQUE \cite{10545385}, these pulses contain some form of obfuscation. Thus, in parallel the encrypted metadata has to be sent to the quantum computer, so it can decrypt it and operate on the input pluses according to the metadata. For example, for \cite{trochatos2023hardwarearchitecturequantumcomputer}, some pulses are attenuated based on the metadata, while for \cite{10545385}, channels on which pulses are supposed to execute are swapped. 
{\em The QOS needs to ensure that the obfuscated circuits of the user are loaded in parallel to the encrypted metadata on the target quantum computer.}
\paragraph{Phase III: Transmission of Computation Results Back to the User}
For each shot of a circuit, it is measured and results returned to the user. Both QC-TEE \cite{10288020} and SoteriaQ \cite{trochatos2023hardwarearchitecturequantumcomputer} proposed to randomly insert {\tt X} gates at the end of the circuit to randomize the output. In parallel, the modified quantum computer hardware generates (and encrypts) its own metadata that can be used by the users to know which qubits' outputs were flipped by the {\tt X}, so the users can recover the correct output. To support these operations, the QOS needs to keep track of the (encrypted) output metadata and transmit it back to the user along with circuit outputs. The transmission back to the user should use secure, classical~networking.
{\em The QOS needs to ensure that the circuit outputs and the output metadata are associated until they are returned to the user. The QOS needs to support classical, secure networking to send back users' results and their encrypted output~metadata.}
\section{Analysis of QOS Support for QTEE}
The QOS modifications to support QTEE are minimal, and can be realized with no overhead on the computation (beyond the overheads of the specific QTEE hardware). Scheduling will be impacted by the QTEEs need that the circuit protection is specific to each quantum computer (because of the unique cryptographic keys needed for the encrypted metadata). QOS scheduler cannot move a circuit to a different quantum computer, since each circuit targets a specific backend. This is not a problem in the current NISQ era, as all circuits are transpiled to a specific back end. But in error corrected quantum computers, where a circuit can execute on different quantum computer backends, this will be a new constraint that the QOS needs to manage.
\section{Conclusion}
The role of the Quantum Operating System (QOS) in supporting Quantum Execution Environments (QTEEs) is crucial to the security of cloud-based quantum computing. By facilitating secure loading of quantum circuits, execution on the quantum computing hardware, and  transmission of computation results back to the users, the QOS can enable robust protection for sensitive quantum computations without compromising performance.
\section*{Acknowledgements}
This work was supported in part by NSF grant \nsf{2245344}.
\bibliography{main}
\bibliographystyle{plain}
\end{document}